\documentstyle [12pt] {article}

\catcode`@=12
\begin{document}
\thispagestyle{empty}
\begin{flushright} UCRHEP-T202\\September 1997\
\end{flushright}
\vspace{0.5in}
\begin{center}
{\Large	\bf Gauged $B - 3 L_\tau$ and Radiative Neutrino Masses\\}
\vspace{1.5in}
{\bf Ernest Ma\\}
\vspace{0.2in}
{\sl Department of Physics, University of California\\}
{\sl Riverside, California 92521, USA\\}
\vspace{1.5in}
\end{center}
\begin{abstract}\
If the minimal standard model of quarks and leptons is extended to include 
just a righthanded partner to $\nu_\tau$, then the quantum number 
$B - 3 L_\tau$ can be added as a gauge symmetry without the appearance of 
anomalies.  A suitable extension of the scalar sector allows one neutrino 
to have a seesaw mass, and the other two to have radiative masses, with 
acceptable phenomenological values for neutrino oscillations.  The $B - 
3 L_\tau$ gauge boson may be light and be observable through its decay into 
$\tau^+ \tau^-$.
\end{abstract}

\newpage
\baselineskip 24pt

The minimal standard model of quarks and leptons possesses four global 
symmetries, {\it i.e.} baryon number ($B$) and the three lepton numbers 
($L_e$, $L_\mu$, $L_\tau$).  They are all conserved at the classical 
level, but are all violated at the quantum level because of the well-known 
axial-vector-current triangle anomaly.\cite{1}  Hence they cannot be 
individually gauged, {\it i.e.} promoted from a global to a local symmetry.  
The linear combination $B - L_e - L_\mu - L_\tau$ 
is a divergenceless current, but it cannot be gauged because the 
corresponding U(1) is anomalous itself without the addition of three 
right-handed neutrino singlets.  In fact, only one of the three 
lepton number differences ($L_e - L_\mu$, $L_e - L_\tau$, $L_\mu - L_\tau$)
can be gauged, leading to some possible interesting phenomenological 
consequences.\cite{2}

If three right-handed neutrino singlets are added, then in general there 
will be mixing among the three lepton families.  Hence it is more 
appropriate to consider a single lepton number $L$ and it is well-known 
that $B - L$ can now be gauged.  In fact, the standard electroweak gauge group 
$SU(2)_L \times U(1)_Y$ is embedded naturally in the left-right 
symmetric gauge group $SU(2)_L \times SU(2)_R \times U(1)_{B-L}$.

In this paper, I will consider instead the addition of just one right-handed 
neutrino singlet and pair it with $\nu_\tau$.  It will be shown that the 
quantum number $B - 3 L_\tau$ can now be added as a gauge symmetry without the 
appearance of anomalies.  The possible existence of the associated gauge 
boson, call it $X$, is very interesting phenomenologically because the 
$B - 3 L_\tau$ gauge symmetry may be broken at an energy scale \underline 
{below} that of electroweak symmetry breaking and yet remains \underline 
{undiscovered}.  Although $X$ shares many of the properties of a gauge boson 
coupled only to $B$, as already discussed in the literature,\cite{3} it will 
be much easier for experiments to confirm or refute because it also couples to 
$\tau$ and $\nu_\tau$.

This model also has important implications regarding neutrinos.  The $\tau$ 
neutrino may acquire a seesaw mass\cite{4} of order 1 eV or 0.1 eV.  The other 
two neutrinos are massless at tree level, but if they mix with $\nu_\tau$ 
through a scalar doublet carrying $L_\tau$ number, then they will acquire 
radiative masses.\cite{5}  A suitable implementation of the scalar 
sector\cite{6} allows realistic values of the masses and mixings of the three 
neutrinos which are consistent with the present experimental evidence for 
neutrino oscillations.
\vspace{0.2in}

\noindent {\bf Cancellation of Anomalies}

Consider the extension of the standard gauge group of particle interactions 
to $SU(3)_C \times SU(2)_L \times U(1)_Y \times U(1)_X$.  Let the quarks and 
leptons transform as follows.
\begin{equation}
\left( \begin{array} {c} u_i \\ d_i \end{array} \right)_L \sim (3, 2, 
1/6; x), ~~~ u_{iR} \sim (3, 1, 2/3; x), ~~~ d_{iR} \sim 
(3, 1, -1/3; x);
\end{equation}
\begin{equation}
\left( \begin{array} {c} \nu_e \\ e \end{array} \right)_L, \left( 
\begin{array} {c} \nu_\mu \\ \mu \end{array} \right)_L \sim (1, 2, 
-1/2; 0), ~~~ e_R, \mu_R \sim (1, 1, -1; 0);
\end{equation}
\begin{equation}
\left( \begin{array} {c} \nu_\tau \\ \tau \end{array} \right)_L \sim 
(1, 2, -1/2; x'), ~~~ \tau_R \sim (1, 1, -1; x'), ~~~ \nu_{\tau R} 
\sim (1, 1, 0; x').
\end{equation}
In the above, only one right-handed neutrino singlet, {\it i.e} 
$\nu_{\tau R}$, has been added to the minimal standard model.  Since the 
number of $SU(2)_L$ doublets remains even (it is in fact unchanged), the 
global SU(2) chiral gauge anomaly\cite{7} is absent.  Since the quarks and 
leptons are chosen to transform vectorially under the new $U(1)_X$, the 
mixed gravitational-gauge anomaly\cite{8} is also absent.  To ensure the 
absence of the axial-vector anomaly,\cite{1} the following conditions are 
considered.\cite{9}

The $[SU(3)]^2 U(1)_X$ and $[U(1)_X]^3$ anomalies are automatically zero 
because of the vectorial nature of $SU(3)$ and $U(1)_X$.  It is also 
easy to show that the $[U(1)_X]^2 U(1)_Y$ anomaly is zero independent 
of $x$ and $x'$.  The remaining two conditions are:
\begin{equation}
[SU(2)]^2 U(1)_X: ~~~ (3) (3) x + x' = 0;
\end{equation}
and
\begin{equation}
[U(1)_Y]^2 U(1)_X: ~~~ (3) (3) [ 2 (1/6)^2 - 
(2/3)^2 - (-1/3)^2 ] x + 
[ 2 (-1/2)^2 - (-1)^2 ] x' = 0.
\end{equation}
Both have the solution: $x' = -9x$.  Let $x = 1/3$ and $x' = -3$, then 
the addition of $U(1)_X$ is recognized as the gauging of $B - 3 L_\tau$.
\vspace{0.2in}

\noindent {\bf Properties of $X$}

Since $X$ does not couple to $e$ or $\mu$ or their corresponding neutrinos, 
there is no direct phenomenological constraint from the best known 
high-energy physics experiments, such as $e^+ e^-$ annihilation, 
deep-inelastic scattering of $e$ or $\mu$ or $\nu_\mu$ on nuclei, or 
the observation of $e^+ e^-$ or $\mu^+ \mu^-$ pairs in hadronic collisions. 
Although $X$ does contribute to purely hadronic interactions, its presence is 
effectively masked by the enormous background due to quantum chromodynamics 
(QCD).  However, unlike the case of a gauge boson coupled only to baryon 
number,\cite{3} $X$ also couples to $L_\tau$.  Assuming that $\nu_{\tau R}$ 
and the $t$ quark are too heavy to be decay products of $X$, the branching 
fraction of $X \rightarrow \tau^+ \tau^-$ is roughly given by
\begin{equation}
B (X \rightarrow \tau^+ \tau^-) = {{(2)(-3)^2} \over {(3)(-3)^2 + (5)(3)(2)
(1/3)^2}} = {54 \over 91}.
\end{equation}
Thus $X$ may be produced in hadronic collisions and be observed through its 
$\tau^+ \tau^-$ signature.
\vspace{0.2in}

\noindent {\bf Scalar Sector}

The minimal scalar content of this model consists of just the usual doublet
\begin{equation}
\left( \begin{array} {c} \phi^+ \\ \phi^0 \end{array} \right) \sim 
(1, 2, 1/2; 0)
\end{equation}
and a neutral singlet
\begin{equation}
\chi^0 \sim (1, 1, 0; 6)
\end{equation}
which couples to $\nu_{\tau R} \nu_{\tau R}$.  As the former acquires a nonzero 
vacuum expectation value, the electroweak gauge symmetry $SU(2)_L \times 
U(1)_Y$ breaks down to $U(1)_{em}$, whereas $\langle \chi^0 \rangle \neq 0$ 
breaks $U(1)_X$.  The resulting theory allows $\nu_{\tau L}$ to obtain 
a seesaw mass and retains $B$ as an additively conserved quantum number and 
$L_\tau$ as a multiplicatively conserved quantum number.  The two other 
neutrinos, {\it i.e.} $\nu_e$ and $\nu_\mu$, are massless in this minimal 
scenario and cannot mix with $\nu_\tau$.

To allow $\nu_e$ and $\nu_\mu$ to become massive without introducing two 
additional right-handed neutrino singlets, the scalar sector is now 
extended to include a doublet
\begin{equation}
\left( \begin{array} {c} \eta^+ \\ \eta^0 \end{array} \right) \sim 
(1, 2, 1/2; -3)
\end{equation}
and a charged singlet
\begin{equation}
\chi^- \sim (1, 1, -1; -3).
\end{equation}
The doublet allows mixing among all three charged leptons and pairs 
$\nu_{\tau R}$ with one linear combination of the three left-handed 
neutrinos.  It appears at first sight that there are then two massless 
neutrinos left.  However, since the three lepton numbers are no longer 
individually conserved, these two neutrinos necessarily pick up 
radiative masses.  This generally happens in two loops through double $W$ 
exchange,\cite{10} but the masses so obtained are extremely small.

To obtain phenomenologically interesting radiative neutrino masses, the 
singlet $\chi^-$ has been added so that there can be the following new 
interactions:
\begin{equation}
f_l (\nu_l \tau_L - l_L \nu_\tau) \chi^+, ~~~ (\phi^+ \eta^0 - \phi^0 \eta^+) 
\chi^- \chi^0,
\end{equation}
where $l = e, \mu$.  The mass-generating radiative mechanisim of Ref.~[6] 
is now operative.  See Figure 1.

One should note that the above scalar sector contains a pseudo-Goldstone 
boson which comes about because there are 3 global U(1) symmetries in the 
Higgs potential and only 2 local U(1) symmetries which get broken.  However, 
if an extra neutral scalar transforming as $(1, 1, 0; -3)$ is added, then 
the Higgs potential will have two more terms and the extra unwanted U(1) 
symmetry is eliminated.
\vspace{0.2in}

\noindent {\bf Neutrino Masses and Mixing}

From the Yukawa couplings $\bar l_L l_R \phi^0$, $\bar \tau_L \tau_R \phi^0$, 
and $\bar \tau_L l_R \eta^0$, the charged-lepton mass matrix linking 
$\bar e_L$, $\bar \mu_L$, $\bar \tau_L$ to $e_R$, $\mu_R$, $\tau_R$ can be 
chosen to be of the form
\begin{equation}
{\cal M}_l = \left[ \begin{array} {c@{\quad}c@{\quad}c} m_e & 0 & 0 \\ 
0 & m_\mu & 0 \\ a_e & a_\mu & m_\tau \end{array} \right].
\end{equation}
The corresponding neutrino mass matrix spanning $\nu_e$, $\nu_\mu$, 
$\nu_\tau$, and $\nu_R^c$ is then given at tree level by
\begin{equation}
{\cal M}_\nu = \left[ \begin{array} {c@{\quad}c@{\quad}c@{\quad}c} 
0 & 0 & 0 & m_1 \\ 0 & 0 & 0 & m_2 \\ 0 & 0 & 0 & m_3 \\ m_1 & m_2 & m_3 & 
M \end{array} \right],
\end{equation}
where $m_{1,2}$ come from $\bar \nu_l \nu_{\tau R} \bar \eta^0$, $m_3$ 
comes from $\bar \nu_\tau \nu_{\tau R} \bar \phi^0$, and $M$ from 
$\nu_{\tau R} \nu_{\tau R} \chi^0$.  Assuming $m_{1,2,3} 
<< M$, the reduced $3 \times 3$ mass matrix spanning the 3 light neutrinos 
becomes
\begin{equation}
{\cal M}_\nu = \left[ \begin{array} {c@{\quad}c@{\quad}c} m_1^2/M & 
m_1 m_2/M & m_1 m_3/M \\ m_1 m_2/M & m_2^2/M & m_2 m_3/M \\ m_1 m_3/M & 
m_2 m_3/M & m_3^2/M \end{array} \right].
\end{equation}
At this stage, there is exactly one neutrino with mass $(m_1^2 + m_2^2 
+ m_3^2)/M$, but there are still 2 massless neutrinos.  Now consider the 
one-loop diagram of Fig.~1.  Of the 3 charged scalars, the linear combination 
$\phi^\pm \cos \theta + \eta^\pm \sin \theta$ (where $\tan \theta = \langle 
\eta^0 \rangle / \langle \phi^0 \rangle$) becomes the longitudinal 
component of $W^\pm$.  The orthogonal combination is physical and mixes 
with $\chi^\pm$.  Let $M_{1,2}$ be the resulting mass eigenvalues and 
$\alpha$ be the mixing angle.  The radiative mass is then easily 
calculated\cite{11} to be
\begin{equation}
m_{\mu \mu} = {{f_\mu a_\mu m_\mu \sin \theta} \over {16 \pi^2 \langle 
\phi^0 \rangle}} \sin \alpha \cos \alpha [F(M_1, m_\tau) - F(M_2, m_\tau)],
\end{equation}
where the function $F$ is given by
\begin{equation}
F(a,b) = {{a^2 \ln (a^2/b^2)} \over {a^2 - b^2}}.
\end{equation}
To illustrate, let $f_\mu = 0.6$, $a_\mu = 10$ MeV, $\sin \theta = 0.01$, 
and $\sin \alpha \cos \alpha [F(M_1, m_\tau) - F(M_2, m_\tau)]$ = 0.01, 
then $m_{\mu \mu} = 2.3 \times 10^{-3}$ eV, which is suitable for 
solar neutrino oscillations\cite{12}.  The other entries of the radiative 
$(\nu_e, \nu_\mu)$ mass matrix are
\begin{equation}
m_{ee} = \left( {{f_e a_e m_e} \over {f_\mu a_\mu m_\mu}} \right) m_{\mu \mu}
\end{equation}
and
\begin{equation}
m_{e \mu} = m_{\mu e} = \left( {{f_e a_\mu m_\mu + f_\mu a_e m_e} \over 
{2 f_\mu a_\mu m_\mu}} \right) m_{\mu \mu}.
\end{equation}
It is clear that the hierarchy $m_{ee} << m_{e \mu} = m_{\mu e} << 
m_{\mu \mu}$ may be naturally established and phenomenologically realistic 
$\nu_e - \nu_\mu$ mixing can be obtained for solar neutrino oscillations. 
If the tree-level neutrino mass of Eq.~(14) is assumed to be of order 1 eV, 
then the mixing of $\nu_\mu$ with $\nu_\tau$ can induce neutrino oscillations 
between $\nu_e$ and $\nu_\mu$ governed by $\Delta m^2 \sim 1$ eV$^2$ as a 
possible explanation\cite{13} of the LSND observations.\cite{14}  If it 
is assumed to be of order 0.1 eV, then atmospheric neutrino 
oscillations\cite{15} may be explained.  Note that $\nu_\tau$-quark 
interactions may affect neutrino oscillations 
inside the sun and the earth, and be a potential explanation of the 
zenith-angle dependence of the atmospheric neutrino deficit.\cite{16}

Because of Eq.~(12), there is also some small mixing of $e, \mu$ into the 
$\tau$ sector.  However, as long as $a_e, a_\mu << m_\tau$, this has 
negligible effects on the current phenomenology.  Details will be presented 
elsewhere.
\vspace{0.2in}

\noindent {\bf $Z - X$ Mixing}

Let $\langle \phi^0 \rangle = v \cos \theta$, $\langle \eta^0 \rangle = v 
\sin \theta$, and $\langle \chi^0 \rangle = u$, then the mass-squared matrix 
spanning $Z$ and $X$ is given by
\begin{equation}
{\cal M}^2_{ZX} = \left[ \begin{array} {c@{\quad}c} (1/2) g_Z^2 v^2 & 
3 g_Z g_X v^2 \sin^2 \theta \\ 3 g_Z g_X v^2 \sin^2 \theta & 18 g_X^2 
(4 u^2 + v^2 \sin^2 \theta) \end{array} \right].
\end{equation}
Precision measurements of the observed $Z$ require this mixing to be very 
small.  Assuming the value $\sin \theta = 0.01$ which was used earlier in 
estimating radiative neutrino masses, this mixing is at most of order 
$10^{-3}$, and that is certainly allowed by the present data.  Note that 
for small $\sin \theta$, $M_X < M_Z$ is indeed possible.  As mentioned 
earlier, this model can be tested by looking for the hadronic production 
of $X$ and observing the decay of $X$ into a $\tau^+ \tau^-$ pair.

It should be noted that there is also mixing of the two U(1) gauge factors 
({\it i.e.} $U(1)_Y$ and $U(1)_X$) in general through the kinetic energy 
terms.\cite{17}  This mixing is assumed to be small here as well.
\vspace{0.2in}

\noindent {\bf Decay of $Z$ to $X$}

If $M_X < M_Z$, then the decay of $Z$ to $X$ and $\bar q q$ for example 
would be possible.  If $X$ couples only to $B$, this would result in four 
hadronic jets, and as shown in Ref.~[3], it would be difficult to discover 
experimentally against the very large QCD background.  However, $X$ also 
couples to $L_\tau$, hence there are the following interesting 
possibilities:
\begin{eqnarray}
(1) &~& Z \rightarrow \bar q q X \rightarrow \bar q q \tau^+ \tau^-, \\ 
(2) &~& Z \rightarrow \bar q q X \rightarrow \bar q q \bar \nu_\tau \nu_\tau, 
\\ (3) &~& Z \rightarrow \tau^+ \tau^- X \rightarrow \tau^+ \tau^- \bar q q, 
\\ (4) &~& Z \rightarrow \tau^+ \tau^- X \rightarrow \tau^+ \tau^- \tau^+ 
\tau^-, \\ (5) &~& Z \rightarrow \tau^+ \tau^- X \rightarrow \tau^+ \tau^- 
\bar \nu_\tau \nu_\tau, \\ (6) &~& Z \rightarrow \bar \nu_\tau \nu_\tau X 
\rightarrow \bar \nu_\tau \nu_\tau \bar q q, \\ (7) &~& Z \rightarrow 
\bar \nu_\tau \nu_\tau X \rightarrow \bar \nu_\tau \nu_\tau \tau^+ \tau^-.
\end{eqnarray}
Thus many distinct signatures are available, such as 2 jets + 2 charged leptons 
+ missing energy from (1) and (3); 2 jets + missing energy from (2) and (6); 
2 charged leptons + missing energy from (5) and (7); and 4 charged leptons + 
missing energy from (4).  Some of the above final states are already being 
searched for in connection with processes such as $Z \rightarrow \bar q q H$, 
where $H$ is a neutral Higgs boson which may decay into $\tau^+ \tau^-$, 
and $Z \rightarrow h A$, where $h$ is a scalar and $A$ a pseudoscalar boson, 
and both decay into $\tau^+ \tau^-$.  However, since the decay of Higgs bosons 
into neutrinos is essentially zero, processes (2) and (5) are unique to this 
model.\cite{18}  In (2), one looks for 2 hadronic jets recoiling against 
nothing, but the missing mass has a sharp peak at $M_X$.  Dedicated searches 
of the above 7 processes will put limits on $M_X$ as well as $g_X$ or discover 
$X$.  More details of the phenomenology will be presented elsewhere.
\vspace{0.2in}

\noindent {\bf Summary and Conclusion}

In this paper, it is shown for the first time that the addition of just one 
right-handed neutrino singlet to the minimal standard model allows the 
quantum number $B - 3 L_\tau$ to be gauged.  This is a novel possibility 
which has hitherto been unrecognized.  Analogously, the associated $X$ 
gauge boson may actually be light and yet remains undiscovered.  However, 
unlike a gauge boson which couples only to baryon number, $X$ decays also 
to $\tau^+ \tau^-$, and that will allow it to be observed in the future 
if it is produced.  Meanwhile, if $M_X < M_Z$, one should look for the decay 
of $Z$ to $X$ in the 7 processes given in Eqs.~(20) to (26).  Unique signatures 
among them are $Z \rightarrow \bar q q X$ and $Z \rightarrow \tau^+ \tau^- X$, 
with $X \rightarrow \bar \nu_\tau \nu_\tau$, {\it i.e.} nothing.

The spontaneous breaking of $B - 3 L_\tau$ through a scalar singlet $\chi^0$ 
retains $B$ as an additively conserved quantum number and $L$ as a 
multiplicatively conserved quantum number.  The addition of a scalar doublet 
and a charged scalar singlet carrying $L_\tau$ number allows the 3 light 
neutrinos to become massive, one by the seesaw mechanism and the other two 
radiatively with possible mixing among all.  Acceptable phenomenological 
values for neutrino oscillations can be obtained.

\noindent {\bf Note added}: After the completion of this paper, it became 
known to the author that the $K^+ \rightarrow \pi^+ \nu \bar \nu$ branching 
fraction has just recently been measured:\cite{19}
\begin{equation}
B = 4.2 \begin{array} {c} +9.7 \\ -3.5 \end{array} \times 10^{-10}.
\end{equation}
This may be a manifestation of the contribution of $X$ to the one-loop 
$W$-induced process $s \rightarrow d \nu_\tau \bar \nu_\tau$.  Details 
will be presented elsewhere.
\vspace{0.3in}
\begin{center} {ACKNOWLEDGEMENT}
\end{center}

I thank D.~P.~Roy for discussions. This work was supported in part by the 
U.~S.~ Department of Energy under Grant No. DE-FG03-94ER40837.

\bibliographystyle{unsrt}

\begin{thebibliography}{99}
\bibitem{18} Actually, a Higgs boson may also decay invisibly, into Majorons 
for example.  However, there should be no confusion because the ratio of the 
$X \rightarrow \bar \nu_\tau \nu_\tau$ 
to $X \rightarrow \tau^+ \tau^-$ widths is predicted here to be exactly 1/2.
\end{thebibliography}

\vspace{0.3in}
\begin{center} {FIGURE CAPTION}
\end{center}

\noindent Fig.~1.  Diagram for generating the one-loop diagonal radiative 
mass for $\nu_\mu$.  Similar diagrams are operative for the rest of the 
$(\nu_e, \nu_\mu)$ mass matrix.

\end{document}